\documentclass[10pt]{article}
\pdfoutput=1
\usepackage{jheppub}

\usepackage{slashed}
\usepackage{color, verbatim}
\usepackage{latexsym}
\usepackage{epsfig}
\usepackage{amsmath,accents}
\usepackage{mathrsfs}

\usepackage{amssymb}
\usepackage{graphicx}
\usepackage{bm}
\usepackage{stackrel}

\usepackage[font={small}]{caption}

\usepackage{hyperref}
\usepackage{epstopdf}
\usepackage{cancel}

\definecolor{myred}{rgb}{0.7, 0, 0}
\definecolor{myblue}{rgb}{0, 0, 0.7}
\definecolor{mygreen}{rgb}{0.04, 0.7, 0.5}
\hypersetup{colorlinks,citecolor=myred,linkcolor=myblue,urlcolor=myblue,linktocpage=true}

\makeatletter

\@addtoreset{equation}{section}
\makeatother

\def \bse {\begin{subequations} \begin{eqnarray}}
\def \ese {\end{eqnarray} \end{subequations}}
\def \be {\begin{equation}}
\def \ee {\end{equation}}
\def \bi {\begin{enumerate}}
\def \ei {\end{enumerate}}
\def \ba {\begin{aligned}}
\def \ea {\end{aligned}}
\def \bse {\begin{subequations} \begin{eqnarray}}
\def \ese {\end{eqnarray} \end{subequations}}

\def \Mp {M_\text{pl}}

\begin{document}

\thispagestyle{empty}

\begin{center}

\vspace{.5cm}

{\LARGE\sc
Baryogenesis from Higgs Inflation
}\\

\vspace{1.cm}

\textsc{\large
Yann Cado, 
Mariano Quir\'os
}\\

\vspace{.5cm}

{\em {Institut de F\'{\i}sica d'Altes Energies (IFAE) and\\ The Barcelona Institute of  Science and Technology (BIST),\\ Campus UAB, 08193 Bellaterra, Barcelona, Spain
}}

\end{center}

\vspace{3cm}

\centerline{\bf Abstract}
\vspace{2 mm}

\begin{quote}

If the inflaton field is coupled to the hypercharge Chern-Simons density $F\tilde F$, an explosive production of helical gauge fields when inflation ends can trigger baryogenesis at the electroweak phase transition. 
Besides, Higgs inflation identifies the inflaton with the Higgs field $\mathcal H$, thus relating cosmological observables to properties of electroweak physics. In this paper we merge both approaches: the helical gauge fields are produced at the end of Higgs inflation from the coupling $|\mathcal H|^2 F\tilde F$. In the metric formulation of gravity we found a window in the parameter space for electroweak baryogenesis consistent with all experimental observations. 
Conversely, for the Palatini formalism the non-gaussianity bounds strongly constrain the helicity produced at the end of inflation, forbidding an efficient baryogenesis.

\end{quote}
 
\vfill

\newpage

\tableofcontents

\newpage

\section{Introduction}
\label{sec:introduction}

The Standard Model (SM) of electroweak and strong interactions of particle physics is a well established theory that, until today, has successfully passed all experimental tests at high-energy colliders (LEP, Tevatron, LHC,...), as well as the low energy ones. Still there are a number of phenomena that cannot be easily coped by the SM, in particular a dynamical explanation of the baryon asymmetry of the universe (BAU), or baryogenesis~\cite{Sakharov:1967dj}, and the existence of cosmological inflation~\cite{Guth:1980zm,Linde:1981mu,Albrecht:1982wi} in the early stages of the universe, both features usually requiring the introduction of beyond the SM (BSM) physics. Still the reluctancy of experimental data to confirm deviations with respect to the SM predictions has motivated people to reanalyze those phenomena with SM tools as much as possible.  

Two main obstacles for the baryogenesis mechanism to work within the SM are: \textit{i)} The required out of equilibrium condition in the electroweak phase transition is forbidden by the present value of the Higgs mass. In fact it has been shown that the electroweak phase transition is not even first order, but a continuous crossover~\cite{Kajantie:1996qd}. \textit{ii)} The CP violation in the CKM matrix is too weak to generate the BAU~\cite{Farrar:1993sp,Farrar:1993hn,Gavela:1993ts,Gavela:1994dt}, so an extra source of CP violation is required. Both problems were solved assuming that inflation is driven by a scalar field, the inflaton, with a dimension 5 operator coupling it to the (CP-odd) hypercharge Chern-Simons density $F\tilde F$ and generating, at the end of inflation, an explosive production of helical hypermagnetic fields~\cite{Anber:2009ua,Anber:2015yca,Cado:2016kdp} relaxing its helicity into the baryon asymmetry at the electroweak crossover~\cite{Kamada:2016eeb,Kamada:2016cnb,Jimenez:2017cdr,Domcke:2019mnd,Cado:2021bia,Cado:2022evn,Cado:2022pxk}.

Cosmological inflation is supposed to be driven by a BSM scalar field with an appropriate potential. The Higgs field $\mathcal H$ with minimal coupling to gravity is excluded as an inflaton candidate, as the quartic coupling is too large to cope with the measured amplitude of density perturbations. It was however observed that if the Higgs field is non-minimally coupled to gravity $\xi_h |\mathcal H|^2R$, with a large coupling $\xi_h$, it can generate cosmological inflation, dubbed as Higgs inflation (HI), consistently with the value of the density perturbations~\cite{Bezrukov:2007ep}. Still HI faces two fundamental problems which possibly require some UV completion of the SM: \textit{i)} Assuming that the Higgs quartic coupling $\lambda_h$ be $\mathcal O(m_h^2/(2v^2))$, the tree level value of the SM, the amplitude of cosmological perturbations~\cite{Planck:2018jri} require that $\xi_h=\mathcal O(10^4)$, which can be at odd with the validity of the effective field theory and violate unitarity constraints. \textit{ii)} When radiative corrections are considered in the SM effective potential, the value of the quartic coupling becomes a function of the Higgs background, $\lambda_h(h)$, which becomes negative, mainly by the contribution from the top quark, at a value $h_I\sim 10^{11}$ GeV~\cite{Degrassi:2012ry}, much below the values for which inflation takes place, i.e.~$h\sim 10^{-2} M_{\rm Pl}$. Problem \textit{i)} has recently been addressed in Refs.~\cite{Antoniadis:2021axu,Ito:2021ssc} where it was proven that, while in the electroweak vacuum tree-level unitarity is violated at the scale $M_{\rm Pl}/\xi_h$, in the inflationary large field background the unitarity limit is at $M_{\rm Pl}/\sqrt{\xi_h}$. Problem \textit{ii)} usually requires an ultraviolet (UV) completion of the model~\cite{Bezrukov:2014ipa} which can modify the relation between the low-energy and high-energy SM parameters, and in particular the value of the coupling $\lambda_h$ at the inflationary scale. Moreover critical HI (CHI) theories~\cite{Bezrukov:2014bra} aim to solve both problems if UV physics modify the running of the SM couplings in such a way that $\lambda_h\ll 1$ at inflationary scales, so that the Higgs mass is near its critical value, while staying positive all the way towards the electroweak scale. For a recent approach in this direction see e.g.~Ref.~\cite{Salvio:2018rv}. In models of CHI it turns out that the amplitude of density perturbations requires values $\xi_h\lesssim 10$ thus greatly alleviating the unitarity problem.

In this paper we will unify both previous approaches and consider cosmological inflation triggered by the Higgs field, i.e.~Higgs inflation, while we will assume that the Higgs is coupled to the hypercharge Chern-Simons density with a CP-odd dimension 6 operator, $|\mathcal H|^2 F\tilde F$. Helical hypermagnetic fields will be generated at the end of inflation, relaxing the helicity into baryon asymmetry at the electroweak crossover. We will then assume ordinary HI with arbitrary value of the quartic coupling, and we will be agnostic about the origin on its value and the mechanism stabilizing the Higgs potential. In this way the value of the coupling $\lambda_h$ during inflation will be considered as a free parameter, which should depend on the particular UV completion of the theory. 

The contents of this paper are as follows. In Sec.~\ref{sec:HI} we review the results on cosmological observables in Higgs inflation. The equations of motion for gauge fields is presented in Sec.~\ref{sec:gauge}, where we prove the (almost) constancy of the parameter $\xi$ which is responsible for the energy transfer from the inflaton to the gauge field, and whose value is the critical quantity for the explosive production of helical  gauge field. The cases of no Schwinger effect (motivated by a solution to the flavor problem by means of a Frogatt-Nielsen mechanism with the flavon coupled to the inflaton) and with Schwinger effect (using a well motivated analytical approximation) are considered in detail. Also the consistency condition for the non-backreaction of the gauge field on the inflaton sector, and the bounds from the non-gaussianity are studied in detail. The baryogenesis capabilities of the model are analyzed in Sec.~\ref{sec:baryogenesis}, both in the cases of Higgs inflation and critical Higgs inflation. We show here that there is an available window where all constrains can be satisfied. As all previous results are done (by default) in the metric formulation of gravity, we have considered in Sec.~\ref{sec:Palatini} the Palatini formulation of gravity, where it is known that the inflationary predictions are different than those in the metric formulation. We have also proven that the baryogenesis predictions are different, and in fact baryogenesis by helical gauge fields is forbidden in the Palatini formulation of gravity. Finally we have drawn our conclusions and outlook in Sec.~\ref{sec:conclusion}, while in App.~\ref{sec:FN} we present the technical details of the Frogatt-Nielsen mechanism where the flavon field is coupled to the inflaton field.

\section{Higgs inflation}
\label{sec:HI}

The model of Higgs inflation is based on an action where the Higgs field  has a non-minimal coupling to gravity. In particular
the action in the Jordan frame is
\begin{equation}
S_J=\int {\rm d}^4x \left[ \sqrt{-g} \left(-\frac{\Mp^2}{2}R-\xi_h |\mathcal H|^2 R +\partial_\mu \mathcal H\partial^\mu \mathcal H^\dag -\dfrac{1}{4} Y_{\mu\nu}Y^{\mu\nu} - U(\mathcal H) \right)-\dfrac{|\mathcal H|^2}{2f_h^2}\;Y_{\mu\nu}\tilde{Y}^{\mu\nu}  \right]
\label{eq:action}
\end{equation}
where $\mathcal H$ is the Higgs doublet, $U$ the Higgs potential in the Jordan frame, and $f_h$ (with mass dimension) provides the inverse coupling of the Higgs to the Chern-Simons term, a CP-odd dimension 6 operator which will be responsible for the baryogenesis mechanism.  $Y^{\mu\nu}$ is the field-strength of the hypercharge gauge field $Y^\mu$, and $\widetilde Y^{\mu\nu}=\frac{1} {2}\epsilon^{\mu\nu\rho\sigma}Y_{\rho\sigma}$ is its dual tensor. A possible UV completion giving rise to this dimension 6 CP-odd operator was provided in Ref.~\cite{Cado:2021bia}.
The action also contains a general non-minimal coupling $\xi_h$ of the Higgs field to the Ricci scalar. 

During the inflationary stage the background physical Higgs field $h$ is large and electroweak (EW) symmetry is broken. The higher-dimensional coupling in Eq.~(\ref{eq:action}) is accordingly replaced by couplings to the photon and the $Z$ boson. Since the $Z$ boson is very massive, we will focus on the coupling to the (massless) photon. Furthermore, all but one degrees of freedom of the Higgs doublet are eaten and we get for the action
\begin{equation}
S_J=  \int {\rm d}^4x \left[ \sqrt{-g} \left(-\dfrac{\Mp^2 + \xi_h h^2}{2} R + \frac{1}{2}\partial_\mu h\partial^\mu h -\dfrac{1}{4} F_{\mu\nu}F^{\mu\nu} - U(h) \right)-\dfrac{\cos^2{\theta_W}}{4}\dfrac{h^2}{f_h^2}\;F_{\mu\nu}\tilde{F}^{\mu\nu}  \right]
\label{eq:actionEinsteinchi-Higgs}
\end{equation}
To alleviate the notation, and for the rest of this paper, we will use units where $\Mp\equiv1$.
For the large values of $h$ involved in inflation, we can use
$ U(h) \simeq \lambda_h h^{4}/4$,
where $\lambda_h$ is taken as a positive parameter. As we have explained in the previous section we are not considering any particular UV completion stabilizing the Higgs potential and, instead, we will consider $\lambda_h$ as a free parameter of the theory. 

To go to the Einstein frame, we perform a Weyl redefinition of the metric
with $ g_{\mu\nu}\to \Theta \, g_{\mu\nu}$ with
\be \Theta(h) =\dfrac{1}{1+ \xi_h h^2} \label{def-theta-mirage}\ee
chosen such that we recover the Einstein-Hilbert action explicitly. 
The potential becomes
\be
V(h)  \simeq \frac{\lambda}{4\xi_h^2}\left(1-\frac{1}{\xi_h h^2} \right)^{2}
\ee
where the approximation is valid in the regime $\xi_h h^2\gg 1$, where the Einstein frame departs from the Jordan frame.
%
%
%

The inflaton field $\chi$, with canonical kinetic term, is related to $h$, by the following change of variables
\be \frac{{\rm d}\chi}{{\rm d}h} = \sqrt{\Theta(1+ 6\xi_h^2 h^2\,\Theta) }, \label{derivative-correction} 
\ee
such that, in the Einstein frame,
\be S_E=  \int {\rm d}^4x\;\sqrt{-g}  \left[ -\dfrac{R}{2} +\frac{1}{2}\partial_\mu \chi\partial^\mu \chi  -\dfrac{1}{4} F_{\mu\nu}F^{\mu\nu} - V[h(\chi)] \right]
-\int d^4x \; \dfrac{\cos^2{\theta_W}}{4}\dfrac{h(\chi)^2}{f_h^2}\;F_{\mu\nu}\tilde{F}^{\mu\nu}  \label{eq:actionEinsteinchi}\ee
where, in the limit $\xi_h h^2\gg1$,  we obtain
\be h(\chi) \simeq  \dfrac{1}{2\sqrt{\xi_h(1+6\xi_h)}} \exp\left(\sqrt{\dfrac{\xi_h}{1+6\xi_h} }\; \chi\right) \label{eq:exact-higgs} \ee
and the potential in terms of the canonically normalized field $\chi$ is then
\be V(\chi) \simeq  \dfrac{\lambda}{4\xi_h^2} \left[ 1-4(1+6\xi_h)\exp\left(-\sqrt{\dfrac{4\xi_h}{1+6\xi_h} }\; \chi\right) \right]^2\,.   \label{eq:exact-pot} \ee
Assuming $\xi_h \gg 1$, which will be proven in the next section, we can write
\be h(\chi) \simeq  \dfrac{e^{\frac{\chi}{\sqrt{6}} } } {\sqrt{24}\,\xi_h}\label{eq:approx-higgs} 
\ee
and
\be  V(\chi)\simeq \dfrac{\lambda}{4\xi_h^2} \left[1-24\xi_h\, e^{- \sqrt{\frac{2}{3}}\;\chi}  \right]^2. \label{pot-approx} \ee

Variation of the action (\ref{eq:actionEinsteinchi}) with respect to $\chi$ and the gauge field $A_\mu= (A_0,\bm{A})$ leads to the gauge equations of motion in the radiation gauge, $A_0=0$ and $\bm{\nabla} \cdot \bm{A} =0$,
\begin{subequations} \label{def:EoM-system} \begin{eqnarray}
 &\ddot{\chi}+3H\dot{\chi}+V'(\chi)= K(\chi)\; \dfrac{\langle\bm{E} \cdot \bm{B}\rangle}{a^4f_\chi}  \label{inflaton-EoM} \\&\left( \dfrac{\partial^2}{\partial \tau^2}-\nabla^2-K(\chi)\;\dfrac{a\, \dot{\chi}}{f_\chi} \; \bm{\nabla} \times \right) \bm{A}= \bm{J} ,\label{gauge-EoM} \ese
 where we are using derivatives with respect to the cosmic time $t$ for the inflaton, as $\dot{\chi}=d\chi/dt$, and with respect to the conformal time $\tau$ for the gauge field, where $d/ d\tau=a d/dt$, and
 \be K(\chi) \equiv \dfrac{{\rm e}^{\sqrt{\frac{2}{3}}\;\chi}}{6^{3/2} \xi_h^2}, \hspace{2cm} f_\chi\equiv\dfrac{2f_h^2}{\cos^2{\theta_W}} .\ee
 Moreover, we have used that $F_{\mu\nu}\tilde{F}^{\mu\nu}=-4\, \bm{E} \cdot \bm{B}$ and $J^\mu=(\rho_c,\bm{J})=ieQ\bar{\psi}\gamma^\mu\psi$, where $\psi$ are the light fermions. 
We assume that initially the Universe does not contain charged particles, and that these ones are produced only later in particle-antiparticle pairs. Therefore, we set the charge density to zero, $\rho_c = 0$. The current $\bm J$ is given by the Ohm's law
$ \bm{J} = \sigma \bm{E} = -  \sigma \partial \bm{A}/\partial \tau$,
where $\sigma$ is the generalized conductivity, to be defined later (see Sec.~\ref{sec:Schwinger-effect}).
%
We assume a homogeneous inflaton 
with only zero-mode, $\chi(t,\bm{x})=\chi(t)$.
All gauge field quantities are $U(1)$ ordinary electromagnetic fields.

Assuming now slow roll inflation, we will neglect the right-hand side of Eq.~(\ref{inflaton-EoM}), i.e.~we will assume $K(\chi) \langle \mathbf{E}\cdot \mathbf{B}\rangle \ll a^4 f_\chi V'(\chi)$, a consistency condition that will be checked a posteriori, after solving the system in Eq.~(\ref{def:EoM-system}). Using the usual slow roll techniques one can easily find the value of the inflaton field at the end of inflation, $\chi_E$, as well as its value $N_\ast$ $e$-folds before, $\chi_\ast$, as 
\begin{align}
\chi_E& = \sqrt{\dfrac{3}{2}} \log \left(24 \xi_h \beta \right) , \hspace{1cm} \beta \equiv 1+\dfrac{2}{\sqrt{3}}\nonumber\\
\chi_\ast &=  \sqrt{\dfrac{3}{2}} \left( \log \left(24 \xi_h \beta \right) -\dfrac{4N_\ast}{3}   -\beta - W_\ast \right)
\label{eq:chiEstar}
\end{align}
where $\mathcal{W_\ast}$ is the Lambert function evaluated at
\be W_\ast \equiv \mathcal{W}_{-1}\left[-\beta \exp{\left(-\beta-\frac{4N_\ast}{3} \right)} \right]. \label{def:chiast}\ee

The slow roll parameters and the cosmic observables can be written as
\be  \epsilon_\ast = \dfrac{4}{3 \left(1+ W_\ast \right)^{2}}, \hspace{2cm}
 \eta_\ast = \dfrac{4\left(2+ W_\ast \right)}{3 \left(1+ W_\ast \right)^2}\,, \ee
so that they are independent on the value of $\xi_h$ and $\lambda_h$. In particular, for $N_\ast = 60\ (50)$ one has
\be \ba \epsilon_\ast &\simeq 0.00019 \ (0.00026)  \hspace{2cm} \eta_\ast \simeq -0.0155 \ (-0.0184) \\ n_s &\simeq 0.968 \ (0.962)  \hspace{2.8cm} r_\ast \simeq 0.003 \ (0.004)\,,
\ea \ee
inside the experimental range measured by Planck~\cite{Planck:2018jri}.

Finally the constraint on the amplitude of scalar fluctuations translates, for $N_\ast=60\, (50)$, into the values for the parameter $\xi_h$
\be
\xi_h\simeq 50\, (42)\cdot 10^3\,\sqrt{\lambda_h}\,,
\label{eq:xilambda}
\ee
which validates our previous  approximation $\xi_h\gg1$ for $\lambda_h \gtrsim 10^{-8}$.

\section{Gauge fields production}
\label{sec:gauge}

 %

We now quantize the gauge field $\bm{A}$ in momentum space
\be \bm{A}(\tau , \bm{x}) \, = \, \sum_{\lambda = \pm} \int \frac{d^3 k}{(2\pi)^3} \, \left [\bm{\epsilon}_\lambda(\bm{k}) \, a_\lambda(\bm{k}) \, A_\lambda(\tau, \bm{k}) \, e^{i \bm{k} \cdot \bm{x}} + \, \text{h.c.} \right]  ,
 \ee
where $\lambda$ is the photon polarization and $a_\lambda(\bm{k})$ ($a_\lambda^\dagger(\bm{k})$) are annihilation (creation) operators that fulfill the canonical commutation relations
$
[a_\lambda(\bm{k}),a_{\lambda'}^\dagger(\bm{k}')]=(2\pi)^3\delta_{\lambda\lambda'}\delta^{(3)}(\bm{k}-\bm{k'})\,.
$
The polarization vectors $\bm{\epsilon}_\lambda(\bm{k})$ satisfy the conditions
\be\begin{aligned} \bm{k} \cdot \bm{\epsilon}_\lambda(\bm{k}) &= 0\, , \hspace{2cm} &   \bm{k} \times \bm{\epsilon}_\lambda(\bm{k})  &= - i \lambda k \, \bm{\epsilon}_\lambda(\bm{k})\, , \\ \bm{\epsilon}^*_{\lambda'}(\bm{k}) \cdot \bm{\epsilon}_\lambda(\bm{k}) &= \delta_{\lambda \lambda'}\, , & \bm{\epsilon}^*_{\lambda}(\bm{k}) &= \bm{\epsilon}_\lambda(-\bm{k}) \, ,\end{aligned}
\label{eq:identitiespol}
\ee
where $k \equiv |\bm{k}|$.
Therefore, the equation of motion for the gauge modes (\ref{gauge-EoM}) yields
\be A''_\lambda+\sigma A'_\lambda  +k \left(k+2\lambda \xi\, aH \right)A_\lambda =0,
\label{eq:Apm}
\ee
where we defined the instability parameter as
\be \xi = -K(\chi)\; \dfrac{\dot{\chi}}{2Hf_\chi}. \label{def:xi} \ee

From the solution of Eq.~(\ref{eq:Apm}), the electric and magnetic energy density, as well as the helicity and helicity time derivative are given by
\begin{subequations} \label{def:EM-quantities} \begin{eqnarray}
\rho_{E} &\equiv& \frac{1}{a^4}  \int^{k_c} dk \,  \frac{k^2}{4 \pi^2}\left(| A'_+|^2 + | A'_-|^2\right), \\
\rho_{B} &\equiv&\frac{1}{a^4} \int^{k_c} d k \,  \frac{k^4}{4 \pi^2}\left(|A_+|^2 + |A_-|^2\right), \\
\mathcal{H} &\equiv& \lim_{V\to\infty}\dfrac{1}{V}\int_Vd^3x \;\frac{\langle \bm{A} \cdot \bm{B} \rangle}{a^3} =\frac{1}{a^3}  \int^{k_c}d k \, \frac{k^3 }{2\pi^2} \left(|A_+|^2-|A_-|^2\right), \label{ABdef} \\
\mathcal{G} &\equiv& \dfrac{1}{2a} \dfrac{d\mathcal{H} }{d\tau}  =-\frac{ \langle \bm{E} \cdot \bm{B}\rangle}{a^4},\end{eqnarray} \end{subequations}
where $A_\pm$ are the solutions of (\ref{eq:Apm}).
The upper integration limit
\be k_c = \left|\frac{a\dot{h}}{2f_h} \right| + \sqrt{  \left(\frac{a\dot{h}}{2f_h} \right)^2+ \frac{a^2}{2}  \left[\dot{\hat{\sigma}}+\hat{\sigma}\left(\frac{\hat{\sigma}}{2}+H\right) \right]},\qquad \hat{\sigma}=\sigma / a \label{cutoff-def} \ee
comes because subhorizon modes have an oscillatory behavior and should be regarded as quantum fluctuations. Therefore, such modes do not contribute to the above classical observables and are excluded from the integration (see Ref.~\cite{Cado:2022pxk} for more details).

\subsection{Almost constancy of $\xi$}
The methods usually employed in the literature to analytically compute the electromagnetic energy density and helicity (at least in the absence of the Schwinger effect) requires a constant $\xi$ (see Refs.~\cite{Anber:2009ua,Anber:2015yca,Cado:2016kdp}. We then aim to demonstrate in this section that this parameter barely changes during inflation while its main dependence lies in the couplings $f_h$ and $\lambda_h$.

First, let us draw our attention on the fact that HI is a good candidate for an (almost) constant $\xi$ as the leading interaction term between the Higgs and the gauge field is a dimension six operator. Let us consider the following interaction term in the action between the inflaton $\phi$ and the Chern-Simons density
\be S_E\supset\int d^4x \;F(\phi)\;Y_{\mu\nu}\tilde{Y}^{\mu\nu}\,.\ee
As the instability parameter is defined by
\be \xi = 2 F'(\phi) \; \frac{V'(\phi)}{V(\phi)} \label{def:xi-F}.\ee
a constant value of $\xi$ is guaranteed by the condition $ F'(\phi) \propto  V(\phi)/V'(\phi)$~\footnote{Note that a dimension 5 operator $\phi F\tilde F$ in a model where the inflaton $\phi$ is an axion-like particle does not provide a constant $\xi$, but rather a exponential behavior given by (\ref{xi-slowroll}) with $K(\chi)=1$, where $\xi$ changes the most at the end of inflation.}. In HI this condition is naturally satisfied as, being $\mathcal H$ an $SU(2)$ doublet, the lowest term in a power expansion is $F(h)\propto h^2$.~\footnote{A linear term in the function $F$ would explicitly break gauge invariance in the symmetric phase.} In this special case we find an (almost) constant value for the parameter $\xi$ during Higgs inflation. In fact, in the slow roll approximation, the instability parameter is given by
\be \xi = \dfrac{K(\chi)}{ f_\chi}  \sqrt{\dfrac{\epsilon}{2}}= \frac{K(\chi)}{f_\chi}\frac{8\sqrt{6}\,\xi_h}{\exp\left[\sqrt{\frac{2}{3}}\chi\right]-24\xi_h}, \label{xi-slowroll} \ee
where, in the second equality, we have used the definition of the potential (\ref{pot-approx}).
Now, using the definition of $K(\chi)$ and $f_\chi$, we obtain that $\xi$ is approximately contant, provided that $\chi \gtrsim  \sqrt{\frac{3}{2}} \Mp \log \left(24 \xi_h\right)\simeq \chi_E$, and given by
\be \xi\simeq \frac{4}{3f_\chi\, \xi_h}\simeq 3.2 \cdot 10^{-5}\; \sqrt{\dfrac{0.1}{\lambda}} \; f_h^{-2}\,.  \label{xi-fit-ana}\ee
 We verified this computation numerically~\footnote{See Ref.~\cite{Cado:2022pxk} for the method.} by solving the full system (\ref{def:EoM-system}) without making the slow roll approximation and found the behavior displayed on the left panel of Fig.~\ref{fig:xi-parameter},
where we plot the parameter $\xi$ as a function of the number of $e$-folds during inflation $N$ for various values of the parameter $f_h$ and $\lambda_h=0.1$. We see in the figure that $\xi$ stays constant during most of the inflationary period and only increases at the end of inflation. The fact that $\xi$ stays almost constant during the inflationary era provides confidence to analytically solve the EoM (\ref{eq:Apm}), while its small variation provides a window for generating baryogenesis, as we will see later on in this paper. 
\begin{figure}[htb]
\begin{center}
\includegraphics[width = 6.8cm]{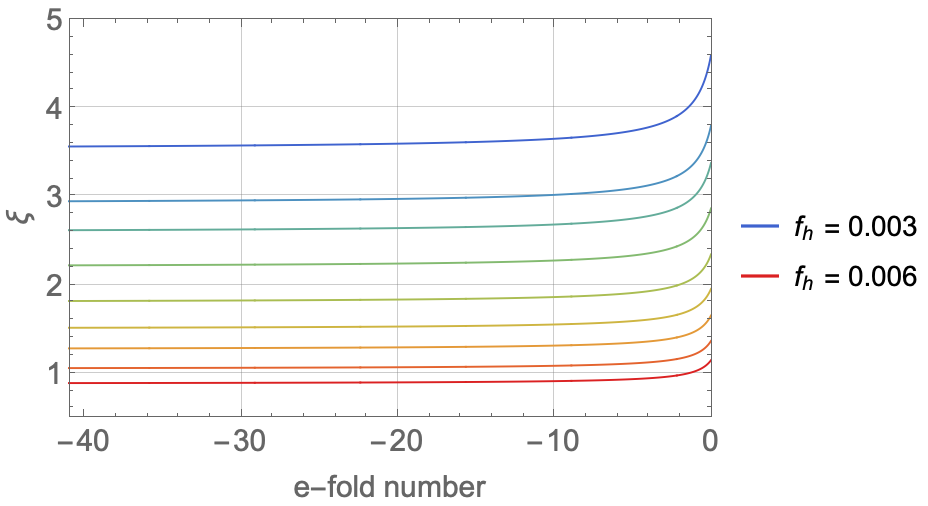}\hspace{1mm}
\includegraphics[width = 8cm]{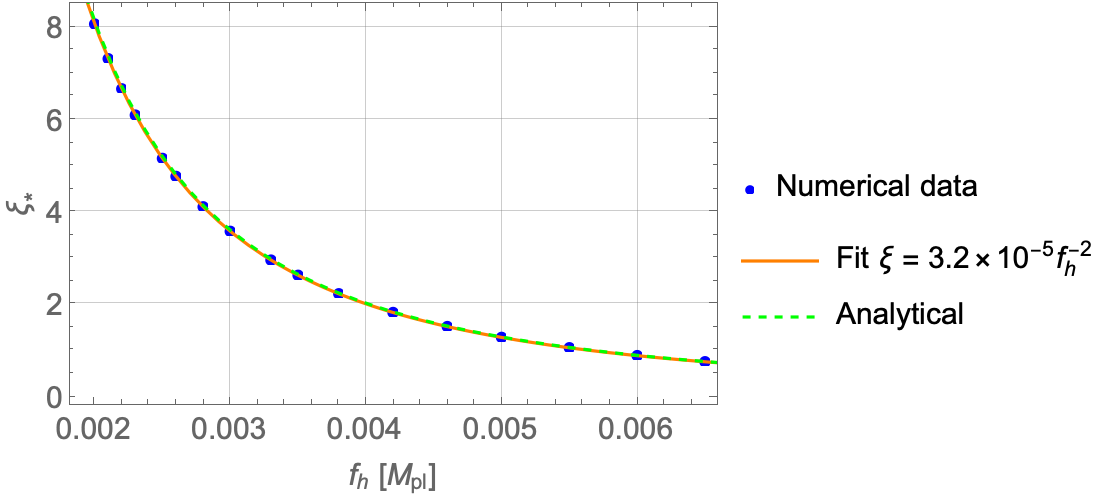}
 \caption{\it Left: Plot of the parameters $\xi$ for various values of the coupling $f_h$.
 Right: Instability parameter at CMB value $\xi_\ast$ for various values of $f_h$ from numerical simulations (blue dots) and their numerical fit (orange). Both perfectly overlap with the analytical relation (\ref{xi-fit-ana}) in dashed green.
 \label{fig:xi-parameter}}
\end{center}
\end{figure}

To know how much $\xi$ does vary during the $N_\ast$ $e$-fold in inflation, we compute, using Eqs.~(\ref{eq:chiEstar}),
\bse \xi(\chi_E) \equiv \xi_E &=& \frac{4}{3f_\chi\, \xi_h}  \frac{\beta}{\beta-1},\\\xi(\chi_\ast) \equiv \xi_\ast &=& \frac{4}{3f_\chi\, \xi_h}  \frac{\beta}{\beta- e^{\frac{4N_\ast}{3}  +\beta + W_\ast}}.\ese
Hence
\be \frac{\xi_E}{\xi_\ast}= \frac{\beta-e^{\frac{4N_\ast}{3}  +\beta + W_\ast}}{\beta-1} \simeq 1.84\,,  \label{eq:xi-ratio}\ee
this ratio being insensitive to the value of $N_\ast$ up the second digit. Notice that it does not contain the self coupling $\lambda_h$, nor $f_h$. 

In conclusion, we see that the instability parameter $\xi$ is flat, regardless on when the simulation begins or on the chosen value of $f_h$. 
Only for the very end of the simulation, $\xi$ deviates from its constant value. In fact, if we plot how this constant value changes with the parameter $\xi_h$, we found a perfect agreement between the numerical calculation and the analytical one (\ref{xi-fit-ana}), as we can see in the right panel of Fig.~\ref{fig:xi-parameter}, where we plot $\xi_\ast$ as a function of $f_h$ for the different estimates.

\subsection{Solution of the gauge equations of motion}
In the presence of strong gauge fields, light fermions charged under the gauge group are produced by the backreaction of gauge fields that source the fermion equations of motion. The corresponding currents can then, in turn, backreact on the produced gauge fields, a phenomenon called \textit{Schwinger effect}. The backreaction of fermion currents on the produced gauge fields acts as a damping force in the explosive production of helical gauge fields. There is nevertheless a condition for a fermion $f$ to contribute to the magnetic conductivity which is, for the fermion Yukawa coupling,
\be
Y_f\lesssim 0.45 \left(\frac{\rho_E}{H^4} \right)^{1/4} \sqrt{|Q_f|}, \label{conductivity-Yukawa-condition}
\ee
and we have computed all couplings at the characteristic scale $\mu\simeq (\langle \bm E\rangle^2+\langle \bm B\rangle^2)^{1/4}$ where the Schwinger effect takes place.
If the three generations of fermions satisfy the above condition then the conductivity for the magnetic field is given by 
\be
\sigma\simeq\frac{e^3}{\pi^2}\frac{a}{H}\;\sqrt{2\rho_B}\;\coth\left(\pi\sqrt{\frac{\rho_B}{\rho_E}} \right), \label{eq:sigma}
\ee
where $e=gg'/\sqrt{g^2+g'^2}$, $e^3\simeq 0.36$.

The case of a constant $\xi$ is suitable for the following scenarios as they both have been studied with this assumption:
\begin{itemize}
\item Absence of the Schwinger effect i.e.~$\sigma \simeq 0$.
\item Presence of the Schwinger effect in the so-called equilibrium estimate~\cite{Domcke:2018eki}.
\end{itemize}
In this section we shall review both cases and compute the baryogenesis parameter space accordingly. 

\subsubsection{No Schwinger effect}
In this section we study the case when the conductivity $\sigma$ vanishes in the equation of motion (\ref{eq:Apm}). One possibility that can guarantee this result would be a dynamical mechanism such that all fermion Yukawa couplings at the inflation scale are $\mathcal O(1)$, such that the criterion (\ref{conductivity-Yukawa-condition}) is not met anymore, and after inflation they relax to the physical values which correspond to fermion masses and mixing angles. A possible mechanism described in App.~\ref{sec:FN} appears if flavor is explained by a 
Frogatt-Nielsen mechanism~\cite{Froggatt:1978nt}, where the flavon field is coupled to the inflaton and get a very large VEV of $\sim h$ during inflation, while the flavon VEV relaxes to its low energy value when $h\simeq v$.


In this setup, we can rewrite (\ref{eq:Apm}) as 
\be A''_\lambda +k \left(k-\lambda \, \frac{2 \xi}{\tau} \right)A_\lambda =0\,,\label{eq:Apm-deSitter-nosigma}
\ee
where we use the scale factor definition $a=-(H \tau)^{-1}$ as we are in de Sitter space.
We solve this equation of motion asymptotically in the slow roll regime. At early time, when $|k\tau|\gg 2\xi $, the modes are in their Bunch-Davies vacuum.
When $|k\tau| \sim 2\xi $, one of the modes develops both parametric and tachyonic instabilities leading to exponential growth while the other stays in the vacuum. During the last $e$-folds of inflation, i.e.~$|k\tau|\ll 2\xi $, the growing mode (wiwth polarization $\lambda$) has the solution~\cite{Anber:2006xt,Anber:2015yca}
\be A_\lambda \simeq \dfrac{1}{\sqrt{2k}}\left(\dfrac{k}{2\xi a_E H_E}\right)^\frac{1}{4} \exp{\left\{\pi \xi-2\sqrt{\dfrac{ 2\xi k}{a_E H_E}}\right\}}, \label{Amplified-mode} \ee
where $a_E$ and $H_E$ are, respectively, the scale factor and the Hubble parameter at the end of inflation. Here, as we assume a slow roll regime, we consider $H_E$ constant and we take the convention $a_E=1$.

Using (\ref{Amplified-mode}), the physical quantities in Eq.~(\ref{def:EM-quantities}) become
\begin{equation}
\rho_{E} \simeq  \dfrac{63}{2^{16}}\; \dfrac{H_E^4}{\pi^2\xi^3}\; {\rm e}^{2\pi \xi} ,  \quad
\rho_{B} \simeq  \dfrac{315}{2^{18}}\; \dfrac{H_E^4}{\pi^2\xi^5}\; {\rm e}^{2\pi \xi} ,  \quad
\mathcal{H}\simeq \dfrac{45}{2^{15}}\; \dfrac{H_E^3}{\pi^2\xi^4} \; {\rm e}^{2\pi \xi}, \quad
\mathcal{G} \simeq \dfrac{135}{2^{16}}\dfrac{H_E^4}{\pi^2\xi^4} \; {\rm e}^{2\pi \xi}.
\label{intQuantities-noSchw}
\end{equation}
In this setup the Hubble can be computed from
$ 3 \Mp^2 H_E^2 \simeq V(\chi_E)$
where the potential is given by Eq.~(\ref{pot-approx}).

These results are only valid when the absence of backreaction on the inflaton EoM (\ref{inflaton-EoM}) is guaranteed, as we will see in Sec.~\ref{sec:consistency}. This model-dependent condition puts a lower bound on the parameter $f_h$ or, equivalently, a higher bound on $\xi$.


\subsubsection{With Schwinger effect \label{sec:Schwinger-effect}}

In cases where the Schwinger effect is at work, we can use the equilibrium Schwinger estimate~\cite{Domcke:2018eki} and redefine $\xi \to \xi_{\rm eff}$ with $\sigma \neq 0$ such that 
\be A''_\lambda  +k \left(k-2\lambda \xi_{\rm eff}\, aH \right)A_\lambda =0.
\ee
with $ \xi_{\rm eff}$ given by the solution of~\cite{Domcke:2018eki}
\be  \,\dfrac{63}{2^{15}\pi^2}\; \dfrac{e^{2\pi \xi_{\rm eff}}}{\xi_{\rm eff}^3} =  \left( \frac{3\pi^2}{e^3}\right)^2 (\xi -\xi_{\rm eff})^2 \;\tanh^2\left(\sqrt{\frac{5}{4}} \,\frac{\pi}{\xi_{\rm eff}} \right). \label{Schw-Equ-Xieff} 
\ee
We show its behavior on Fig.~\ref{fig:xieff} where we plot the effective parameter $\xi_{\rm eff}$ as a function of $\xi$. In this approximation the prediction for the gauge quantities in Eqs.~(\ref{intQuantities-noSchw}) as those in the backreactionless scenario with the replacement $\xi\to \xi_{\rm eff}$.
%
%
\begin{figure}[htb]
\begin{center}
\includegraphics[width = 7cm]{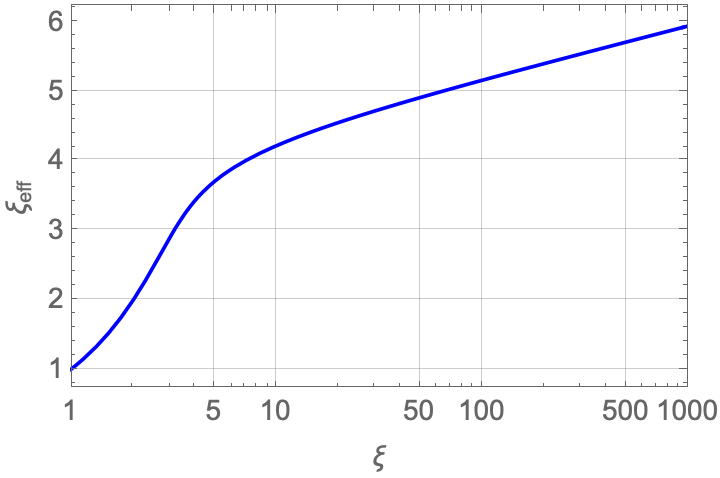}
 \caption{\it In the Schwinger equilibrium estimate, the instability parameter $\xi$ is replaced with an effective one that mimic the fermion backreaction on the gauge fields. We display their relation in the above plot.
 \label{fig:xieff}}
\end{center}
\end{figure}
The consistency condition and the non-gaussianity bounds that we will present, respectively, in Secs.~\ref{sec:consistency} and \ref{sec:non-gaussianity}, should apply to the parameter $\xi$ in the backreactionless case, and to the effective parameter $\xi_{\rm eff}$ in the case of the Schwinger effect with equilibrium solution.

\subsection{Backreactionless consistency condition \label{sec:consistency}}
In the absence of backreaction of the gauge field on the inflaton EoM, the inflationary equation (\ref{inflaton-EoM}) with slow roll conditions reduces to $3H\dot{\chi} \simeq -V'(\chi)$. Thus, in order to consistently neglect the backreaction on the inflaton, we must simply enforce that, in the inflaton EoM (\ref{inflaton-EoM}), the right-hand side term is negligible as compared to the kinetic term, i.e.
\be  3 H \dot{\chi} \gg K(\chi) \, \dfrac{\mathcal{G}}{f_\chi}. \label{noRHS-crit} \ee

Using the result (\ref{intQuantities-noSchw}) for $\mathcal{G}$ and the definition of $\xi$ (\ref{def:xi}), this condition becomes
 \be \frac{45}{2^{13}}\frac{e^{2\pi \xi}}{\xi^3} \ll \mathcal P^{-1}_\zeta
 \ee
 where the spectrum of primordial perturbations, for around 60 $e$-folds before the end of inflation (i.e.~for $\chi=\chi_\ast$) is $\mathcal P_\zeta^{1/2}=H^2/(2\pi |\dot\chi|)\simeq 4.7 \times 10^{-5}$~\cite{Planck:2018jri}. This leads to the upper bound 
$\xi_\ast\lesssim 4.74$, for which we can neglect the backreaction of the gauge fields on the inflaton EoM for the value of the inflaton field $\chi=\chi_\ast$. As we will see in the next section this condition is superseded by the condition of non-gaussianity effects. 

We must however ensure that condition (\ref{noRHS-crit}) is valid throughout the end of inflation. Using the slow roll conditions, and the fact that, for our model, $V'(\chi_E)>V'(\chi_\ast)$, we found a stronger bound than the former one as Eq.~(\ref{noRHS-crit}) can be written as
 \be
 \xi\, \mathcal G\ll \frac{V'^2}{6H^2},
 \ee
 which leads to $\xi_E\lesssim 6.45$ (i.e.~$\xi_\ast \lesssim 3.48$), at the end of inflation.
 
 Once the non backreaction condition on the inflaton equation is satisfied, the no backreaction condition on the Friedmann equation 
 \be
 \frac{\langle \mathbf E^2+\mathbf B^2\rangle}{2a^4}=\frac{63}{2^{16}}\frac{H^4}{\pi^2\xi^3}e^{2\pi\xi}\left(1+\frac{5}{4\xi} \right)\ll V\simeq3 H^2 
 \ee
 holds automatically. In particular the latter condition leads to $\xi_E\lesssim 6.55$  (i.e.~$\xi_\ast \lesssim 3.54$).

\subsection{Non-gaussianity bounds for HI}
\label{sec:non-gaussianity}
As pointed out in Refs.~\cite{Barnaby:2010vf,Barnaby:2011qe}, even if the non-backreaction conditions are satisfied, the coupling $h^2 F \tilde F$ can generate cosmological fluctuations in the HI model. The perturbations on the inflaton are obtained by replacing $\chi(t,\vec x)= \bar\chi(t)+\delta\chi(t,\vec x)$, where $\bar\chi(t)$ is the inflationary background and $\delta\chi(t,\vec x)$ the fluctuation. The equation for the fluctuation is given by
\be \left[ \frac{\partial^2}{\partial t^2}+3H\frac{\partial}{\partial t} - \frac{\nabla^2}{a^2} +V''(\bar{\chi}) - \bar{K}' \frac{\mathcal{G}}{f_\chi} \right]\delta \chi = K(\chi)\;\frac{\delta \mathcal{G}}{f_\chi} \label{eq:perturbations} \ee
where $\bar{K}\equiv K(\bar\chi)$ and $\delta \mathcal {G}=(\mathbf E\cdot \mathbf B-\langle \mathbf E\cdot \mathbf B\rangle)/a^4$.

The function $\bar K$ satisfies the condition $\bar K'=\sqrt{2/3}\bar K$, while for our potential, during the inflationary period, it turns out that $V''(\bar\chi)\simeq -\sqrt{2/3}V'(\bar\chi)$. Then, the last two terms of the left-hand side of Eq.~(\ref{eq:perturbations}), are
\be
V''-\frac{\bar K'}{f_\chi}\mathcal G\simeq -\sqrt{\frac{2}{3}}\left(V'+\frac{\bar K}{f_\chi}\mathcal G \right)\simeq -\sqrt{\frac{2}{3}} V'\simeq V''
\ee
where we have made use of the non-backreaction condition (\ref{noRHS-crit}). In this way the last term in the left-hand side of Eq.~(\ref{eq:perturbations}) can be safely neglected.

The resulting fluctuation equation has been explicitly solved in Ref.~\cite{Barnaby:2011vw}, provided the backreactionless consistency condition of Sec.~\ref{sec:consistency} is satisfied,
%
%
as well as the correlation functions for the curvature perturbations on uniform density hypersurfaces $\zeta(t,\vec x)=-H\delta\chi(t,\vec x)/\dot{\bar\chi}$. A good fit for the equilateral configuration of the three-point function yields the fit, valid for values $2\lesssim \xi\lesssim 3$~\cite{Barnaby:2011vw},
\be
f_{\rm NL}^{\rm equil}\simeq \frac{1.6 \times10^{-16}}{\xi^{8.1}}e^{6\pi\xi}
\ee

The current Planck bound on $f_{\rm NL}^{\rm equil}$~\cite{Planck:2019kim},
$
f_{\rm NL}^{\rm equil}=-26\pm 47
$
yields, at CMB scales, the upper bound $\xi_\ast\lesssim 2.55$, at $95\%$ CL. A  much stronger condition than that leading to the absence of backreaction. 
Given that in our model the almost constancy of $\xi$ leads to the relation (\ref{eq:xi-ratio}) the non-gaussianity bound translates in our model into the bound \fbox{$\xi_E\lesssim 4.71$}.
As already stated, all the calculations done in the absence of Schwinger effect apply, in the presence of Schwinger effect in the equilibrium approximation, to corresponding bounds on the effective parameter, i.e.~$\xi_{{\rm eff}\ast}<2.55$. 

\section{Baryogenesis \label{sec:baryogenesis}}

We will follow in this section the formalism and technical details from Ref.~\cite{Cado:2022evn} for the baryogenesis mechanism. In particular the value of the baryon-to-entropy ratio generated by the decay of the helicity at the electroweak phase transition is given by 
\be\eta_B  \simeq   4 \cdot 10^{-12} \, f_{\theta_W}  \frac{\mathcal{H}}{H_E^3} \left( \frac{H_E}{10^{13} \, \text{GeV}} \right)^{\frac{3}{2}} \left(\frac{T_{\rm rh}}{T_{\rm rh}^{\rm ins}}  \right) \, \, \simeq \, 9 \cdot 10^{-11} ,\label{constraint-nB} \ee
where the last equality is the observational value~\cite{Zyla:2020zbs}. Following Refs.~\cite{Domcke:2019mnd, Cado:2021bia} we define the parameter $f_{\theta_W}$, which encodes all the details of the EW phase transition and its uncertainties, 
\be f_{\theta_W}  = -\sin (2 \theta_W) \, \dfrac{d\theta_W}{d\ln T}\bigg\rvert_{T=135\text{ GeV}}, \hspace{1.5cm} 5.6 \cdot 10^{-4} \lesssim f_{\theta_W}  \lesssim 0.32. 
\label{eq:ftheta}
\ee

We assume instant reheating~\cite{Sfakianakis:2018lzf,Rubio:2019ypq,Dux:2022kuk},  $T_{\rm rh}\simeq T_{\rm rh}^{\rm ins}$,
hence the ratio $T_{\rm rh}/T_{\rm rh}^{\rm ins}$ drops in Eq.~(\ref{constraint-nB}).
However, in addition to their dependence on the gauge sector observables, the quantities used in this section vary according to the quartic coupling $\lambda_h$ as $\xi \propto \lambda_h^{-1/2}$, see Eq.~(\ref{xi-fit-ana}). Besides, the Hubble ratio at the end of inflation 
$ H_E \simeq \sqrt{V(\chi_E)/3}$
also depend on $\lambda_h$ as $V$ does.
\subsection{Constraints}
There are however, a number of constraints that must be fulfilled before any claim on the BAU can be made, see Ref.~\cite{Cado:2022evn}. To ensure that the required magnetohydrodynamical conditions are fulfilled for the (hyper)magnetic fields to survive until the electroweak crossover, we must demand that the magnetic Reynold's number at reheating $\mathcal R_m^{\rm rh}$ is bigger than one.
As we are in the viscous regime, we can compute
\be\mathcal{R}_m^{\rm rh}  \approx 5.9 \cdot 10^{-6} \; \frac{\rho_{B} \ell_{B}^2}{H_E^2}  \left( \frac{H_E}{10^{13} \, \text{GeV}} \right) \left(\frac{T_{\rm rh}}{T_{\rm rh}^{\rm ins}}  \right)^{\frac{2}{3}},\label{constraint-Rm}\ee
where $\ell_{B}$ is the physical correlation length of the magnetic field given by
\be \ell_{B}  =\frac{2\pi}{\rho_B\,a^3}  \int^{k_c}d k \, \frac{k^3 }{4\pi^2} \left(|A_+|^2+|A_-|^2\right)\simeq \dfrac{8}{7} \dfrac{\pi\,\xi}{H_E} ,\ee
where in the second step we use the solution (\ref{Amplified-mode}).

Then, the chiral plasma instability (CPI) temperature must be low enough to ensure that the CPI time scale is long enough to allow all right-handed fermionic states to come into chemical equilibrium with the left-handed ones via Yukawa coupling interactions (so that sphalerons can erase their corresponding asymmetries in particle number densities) before CPI can happen. The estimated temperature at which CPI takes place is
\be T_{\rm CPI}/\textrm{GeV}  \approx    4 \cdot 10^{-7} \,  \; \frac{\mathcal{H}^2}{H_E^6} \, \left( \frac{H_E}{10^{13} \, \text{GeV}} \right)^3\left(\frac{T_{\rm rh}}{T_{\rm rh}^{\rm ins}}  \right)^2 \,.
\label{constraint-TCPI}\ee
The constraint $T_{\rm CPI}\lesssim 10^5$~GeV (the temperature at which $e_R$ comes into chemical equilibrium) guarantees that the CPI cannot occur before the smallest Yukawa coupling reaches equilibrium and all particle number density asymmetries are erased, preventing thus the cancellation of the helicity generated at the reheating temperature.

Finally, with the values of energy densities and helicity at our hand we checked that the generation of baryon isocurvature perturbation provides no constraint.

\subsection{Higgs inflation}
As we have previously explained we will be agnostic about the mechanism stabilizing the Higgs potential and then just will consider $\lambda_h$ as a free parameter. The corresponding plot, for values $10^{-3}\lesssim \lambda_h\lesssim 1$, is shown in Fig.~\ref{fig:BAU}, for the backreactionless case (left panel) and the Schwinger equilibrium solution (right panel), which shows that condition (\ref{constraint-nB}) provides a wide window for baryogenesis (in blue). Then we display in orange the region where  $\mathcal R_m^{\rm rh}>1$, see (\ref{constraint-Rm}), and in green the region where $T_{\rm CPI} \lesssim 10^5$~GeV, see (\ref{constraint-TCPI}). 
In both plots the red region is excluded because of the CMB non-gaussianity bound.
\begin{figure}[htb]
\begin{center}
\includegraphics[width = 8cm]{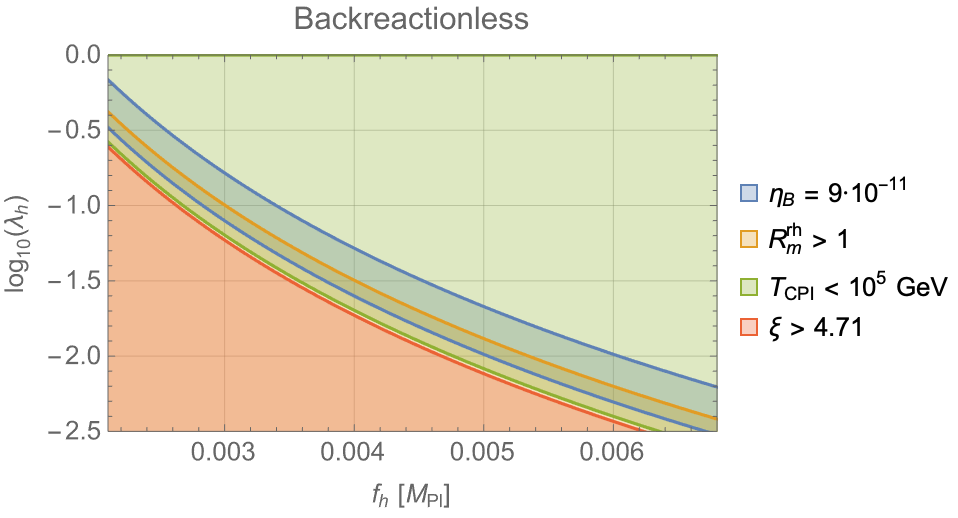}
\includegraphics[width = 6cm]{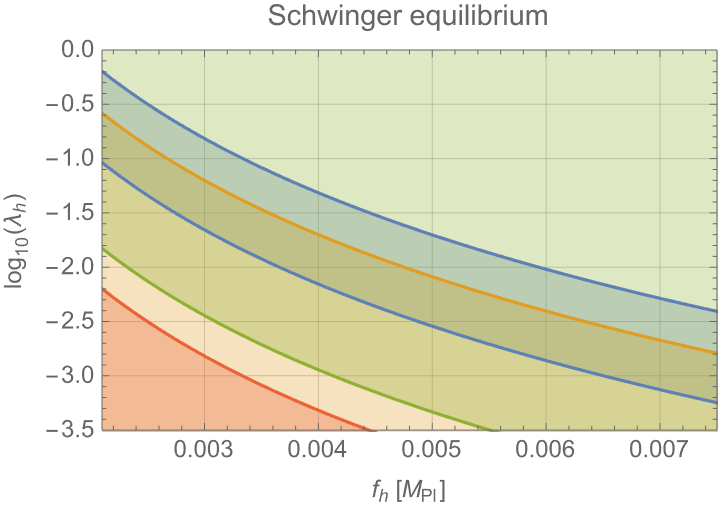}
 \caption{\it The baryogenesis parameter space for the backreactionless (left panel) and Schwinger equilibrium (right panel) cases. The red region is excluded because of CMB non-Gaussianity. We seek the overlapping region between the first three one. The condition on CPI temperature is no constraint since it overlaps the entire region for $\eta_B$. Hence the tradeoff must be made between $\eta_B$ and the magnetic Reynolds number. 
  \label{fig:BAU}}
\end{center}
\end{figure}

We can see that in this scenario, the BAU is attained for values
\be 3.6 \lesssim \xi_E \lesssim 4.1. 
\label{eq:window}
 \ee
This range is the same for both the backreactionless and the Schwinger equilibrium case by construction of the latter. However, because of the replacement $\xi \to \xi_{\rm eff}$, the relation between $\xi$ and the couplings $\lambda$ and $f_h$ is different in both cases (this is why we showed two panels on Fig.~\ref{fig:BAU}).
These bounds correspond to the values
\be \begin{aligned} 1.4 \times 10^4 \lesssim \frac{\rho_E}{H_E^4} \lesssim  1.7\times 10^5 \hspace{2cm}  1.4 \times 10^3 \lesssim \frac{\rho_B}{H_E^4}  \lesssim  1.3 \times 10^4 \\  5.6\times 10^3 \lesssim \frac{\mathcal{H}}{H_E^3} \lesssim  6.2\times 10^4 \hspace{2cm}  8.4 \times 10^3 \lesssim \frac{\mathcal{G}}{H_E^4}  \lesssim  9.3 \times 10^4 \end{aligned} \ee

\subsection{Critical Higgs Inflation \label{sec:CHI}}
Depending on the values of the Higgs and top quark masses,
$\lambda_h$ could remain positive till the Planck scale, and such that $\lambda_h\ll 1$ and $\beta_{\lambda_h}\ll 1$ (exhibiting a \textit{critical behavior}) without any need of new physics. In particular this should happen if the top quark mass is $m_t \simeq 171.3$~GeV~\cite{Degrassi:2012ry,Alekhin:2012py,Buttazzo:2013uya}, which however exceeds its current value from direct measurements, $m_t=172.76\pm 0.30$ ~\cite{ParticleDataGroup:2022pth} by $\sim 3\sigma$. Those models initially proposed in Refs.~\cite{Bezrukov:2014bra,Hamada:2014iga,Hamada:2014wna,Ezquiaga:2017fvi,Salvio:2017oyf,Masina:2018ejw,Drees:2019xpp} were dubbed critical Higgs inflation (CHI) and in principle would not need any UV completion for the Higgs potential stabilization. 

Nevertheless, in view of the actual experimental values of the Higgs and top quark masses, people have been proposing UV completions changing the size of the quartic $\beta$-function, and such that $\lambda_h$, and $\beta_{\lambda_h}$, can attain a critical behavior for the values of the Higgs for which HI takes place, and stay positive all the way down to the electroweak scale~\cite{Salvio:2018rv}. 

In all cases, for critical values of $\lambda_h$, CHI has the advantage that the required value of the coupling to the Ricci scalar $\xi_h$, as given by Eq.~(\ref{eq:xilambda}), is considerably reduced with respect to ordinary HI. In particular $\xi_h\lesssim \mathcal O(10)$ for $\lambda_h\lesssim 4\cdot 10^{-8}$.
\begin{figure}[htb]
\begin{center}
\includegraphics[width = 7cm]{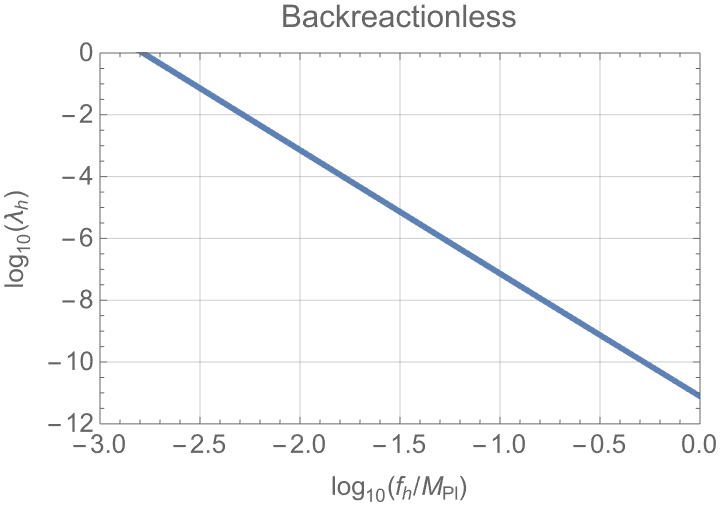}\hspace{1cm}
\includegraphics[width = 7cm]{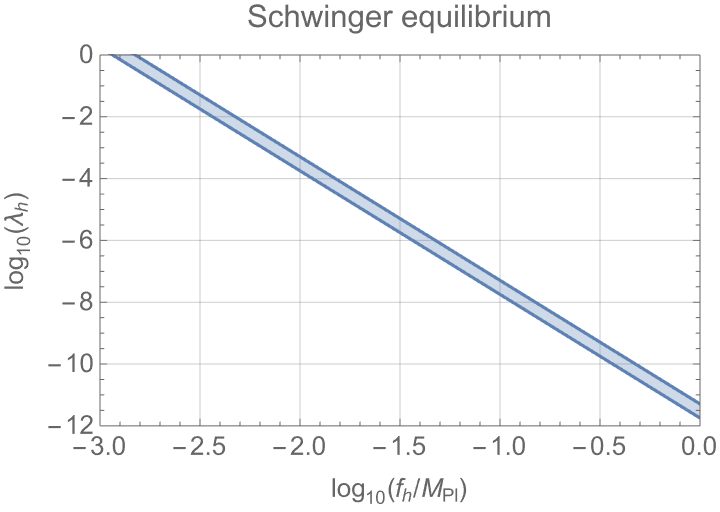}
 \caption{\it Region where the BAU can successfully be achieved, for a wider range of the parameters.
 \label{fig:CHI-baryogenesis}}
\end{center}
\end{figure}
For these reasons, we found it interesting to show a wider parameter window of Fig.~\ref{fig:BAU} that covers smaller values of the self coupling parameter $\lambda_h$.
We show, in Fig.~\ref{fig:CHI-baryogenesis}, the overlapping region of Fig.~\ref{fig:BAU} for $\lambda_h \ll 1$ where all conditions are met to successfully produce the BAU.
As in this case, the Higgs self coupling can be arbitrary small, we used the exact solutions (\ref{eq:exact-higgs}) and (\ref{eq:exact-pot}) instead of their approximations (\ref{eq:approx-higgs}) and (\ref{pot-approx}), with only minor differences.

\section{Palatini formulation}
\label{sec:Palatini}
In this paper we have used the metric formulation of gravity, where the connection giving rise to the Ricci scalar is identified with the Levi-Civita connection 
$\Gamma^{\rho}_{\ \mu\nu}$, and thus related to the metric $g_{\mu\nu}$. There is an alternative formulation, the Palatini formulation of gravity, where the connection is arbitrary and torsion-free, i.e.~$\Gamma^{\rho}_{\ \mu\nu}=\Gamma^{\rho}_{\ \nu\mu}$. One of the main features of the Palatini formalism is that the inflationary predictions are different than those in the metric one~\cite{Bauer:2008zj}.

In the Palatini HI (for a review, see e.g.~Ref.~\cite{Rubio:2019ypq}), the connexion from which the Ricci tensor is calculated does not depend on the metric, and the Weyl rescaling (\ref{def-theta-mirage}) leaves $R$ invariant. Hence, in the Einstein frame, the Palatini action is written as
\be S_E=  \int {\rm d}^4x\;\sqrt{-g}  \left[ -\dfrac{R}{2} + \frac{\Theta^2}{2} \;  \partial_\mu h\partial^\mu h -\dfrac{1}{4} Y_{\mu\nu}Y^{\mu\nu} - V(h) \right]\nonumber  -\int d^4x \;F(h)\;F_{\mu\nu}\tilde{F}^{\mu\nu},
\label{eq:actionEinstein-Palatini}
\ee
where $\Theta$ is given by (\ref{def-theta-mirage}), and the canonical inflaton $\chi$ is obtained by 
\be \frac{{\rm d}\chi}{{\rm d}h} = \sqrt{\Theta} = \dfrac{1}{\sqrt{1+\xi_h h^2}}  .
\ee
This considerably  simplifies the equations in terms of $\chi$ as we can now write exact analytical relations such as
\bse \chi(h)&=&  \dfrac{\sinh\left(\sqrt{\xi_h}\, h\right)}{\sqrt{\xi_h}} ,\\ V(\chi) &=& \dfrac{\lambda}{4\xi_h^2}\tanh^4\left(\sqrt{\xi_h}\,\chi \right) . \ese

The slow roll analysis from HI in the metric formulation is then modified as we now have
\bse   \sinh(2\sqrt{\xi_h}\chi_E)&=&4\sqrt{2\xi_h} \\  \cosh(2\sqrt{\xi_h}\chi_\ast)&\simeq& 16 \xi_h  N_\ast \label{palatini-chiast}, \ese
%
%
 %
Using Eq.~(\ref{palatini-chiast}), the amplitude of scalar fluctuations at $N_\ast$ 
 \be
 A_s
 =
 \frac{\lambda_h}{768\pi^2}\frac{\sinh^4(\sqrt{\xi_h}\chi_\ast)\tanh^2(\sqrt{\xi_h}\chi_\ast)}{\xi_h^3}
 \ee
 %
 leads to
 \be
 \xi_h\simeq 1.4 \times 10^9 \; \frac{\lambda}{0.1} \left(\frac{N_\ast}{60}\right)^2 .
 \ee
 %
 %

Finally, considering the coupling to the Chern-Simons density $F(\chi)\,F\tilde F$ given by the quadratic function
 \be
 F(\chi)=\dfrac{\cos^2{\theta_W}}{4} \frac{h^2(\chi)}{f_h^2}=\dfrac{\cos^2{\theta_W}}{4}\frac{\sinh^2(\sqrt{\xi_h}\chi)}{\xi_h f_h^2}
 \ee
 the parameter $\xi$ is given, using Eq.~(\ref{def:xi-F}), by
 \be
 \xi=\frac{4\cos^2{\theta_W}}{f_h^2}.
 \ee
 which is constant throughout all the inflationary period. 
 
 Notice the important difference between HI in the metric and Palatini formalisms. While in the former the parameter $\xi$ is almost constant, just providing a small growth $\xi_E\sim 1.84\xi_\ast$ at the end of inflation, in the latter the parameter $\xi$ is exactly a constant throughout all the inflationary period. Therefore while in the metric formulation the non-gaussianity bound on $\xi$ at the CMB $\xi_\ast<2.55$ translates into the bound $\xi_E<4.71$ at the end of inflation, when the helical magnetic fields are generated, relaxing its helicity into the baryon asymmetry at the electroweak phase transition, in the Palatini formulation the non-gaussianity bound at the end of inflation is $\xi_E<2.55$. Given the baryogenesis window (\ref{eq:window}) we have found, this result means that, while Palatini HI can be a viable candidate to produce cosmological inflation, however the magnetic fields produced at the end of Palatini HI have not enough strength to generate the baryon asymmetry of the universe. 
 

\section{Conclusion}
\label{sec:conclusion}
Baryogenesis and cosmological inflation are two main issues which usually require the existence of BSM physics. \textit{i)} The baryogenesis mechanism is too weak in the SM for the present values of the Higgs boson, as the electroweak phase transition is too weak (a crossover) and the amount of CP-violation induced by the CKM phase too small due to the presence of light quark masses. Thus most baryogenesis mechanisms rely on BSM extensions for which the electroweak phase transition is strong first order and have an extra source of CP-violation. Still there is a tension between electric dipole moment (EDM) bounds and the required amount of BAU. \textit{ii)} On the other hand, cosmological inflation requires the presence of an extra BSM field $\chi$, the inflaton, with an appropriately flat potential. In view of the lack of experimental evidence for BSM physics at low energy, there have been attempts to solve the above problems with as much as possible SM physics. 

\textit{i)}
Concerning the baryogenesis mechanism, in the presence of the inflaton coupling to the Chern-Simons hypercharge density $\chi F\tilde F$, generating CP-violation, helical gauge fields can be produced at the end of inflation and the helicity relaxes to baryon asymmetry at the electroweak crossover generating the observed BAU. In this way the physics at the electroweak breaking scale is that provided by the SM of electroweak and strong interactions. 

\textit{ii)}
Concerning the problem of cosmological inflation, it was proven that the Higgs field $\mathcal H$ can generate enough inflation, consistent with cosmological observations by the Planck collaboration, provided that it is non-minimally coupled to gravity. In this case one could achieve cosmological inflation with the SM degrees of freedom. Still this approach has some caveats. One of them being that, in the SM, for the current values of the Higgs and top-quark masses, the Higgs self coupling is driven to negative values at scales $\sim 10^{11}$ GeV, much lower than the inflationary (Planckian) scales, so one needs some UV completion to change the RGE evolution of the SM couplings, or perhaps some criticality value of the SM quartic coupling at the inflationary scales. 

In this paper we have merged both above approaches. In particular we have considered Higgs inflation, where the Higgs is non-minimally coupled to gravity, and added a dimension 6 CP-violating operator coupling the Higgs to the hypercharge Chern-Simons density: $|\mathcal H|^2 F \tilde F$. We have proven there is an explosive production of helical hypermagnetic fields which can produce baryogenesis when the helicity relaxes into the BAU at the electroweak crossover. The parameter $\xi$ responsible for the energy transfer from the inflaton to the gauge fields is almost a constant, due to the particular shape of the inflationary potential and the coupling of the Higgs to the Chern-Simons density, and we can thus fully rely on analytic approximations to consider the gauge field solutions. We have also proven that the helicity produced at the end of inflation satisfies the required magnetohydrodynamical conditions to survive to the electroweak phase transition, and produce the observed BAU, for a window of $\xi$ at the CMB scales given by $1.96<\xi_\ast<2.23$ (corresponding at the end of inflation to $3.6<\xi_E<4.1$), thus satisfying the bound $\xi_\ast<2.55$ on non-gaussianity. 

In the above analysis we have worked in the metric formulation of gravity and considered two especially simple cases: \textit{a)} In the absence of Schwinger effect, and; \textit{b)} In the presence of Schwinger effect. We have implemented case \textit{a)} by assuming that the SM flavor problem is implemented by means of a Frogatt-Nielsen mechanism, in the case where the flavon field is coupled to the inflaton. As a consequence of this coupling, during inflation one can easily impose the condition that all fermions be heavy (say as heavy as the top quark) in such a way that the Schwinger conductivity, which is exponentially suppressed by the fermion mass squared, is negligible and the Schwinger effect turns out to also be negligible. After inflation the flavon field relaxes to its usual minimum which can describe all fermion masses and mixing angles at the electroweak scale. The details of the mechanism are described in App.~\ref{sec:FN}. As for case \textit{b)}, in the presence of the Schwinger effect, we have taken advantge of the (almost) constancy of the parameter $\xi$ to use the simple Schwinger equilibrium approximation, which simply amounts to a redefinition of the $\xi$ parameter. In all cases we have extended our calculation to the case of critical Higgs inflation and found that for values of the quartic Higgs self-coupling $\lesssim 10^{-10}$ the coupling $1/f_h$ of the Higgs to the Chern-Simons density $\frac{h^2}{f_h^2}F\tilde F$ can be $\lesssim  M_{\rm Pl}^{-1}$, in the weakly coupled region. 

We also have considered the Palatini formulation of gravity. In this case the equations for the change from the Jordan to the Einstein frame are analytic, as well as the inflationary potential and the relation between the inflaton $\chi$ and the Higgs field $h$. As a consequence of the shape of the inflationary potential it turns out that in this model the parameter $\xi$ is exactly a constant, i.e.~$\xi_\ast=\xi_E$. In this formalism helical gauge fields can be produced, however the bounds on non-gaussianity imposes that its production is not so explosive as required to trigger electroweak baryogenesis, which is then forbidden in this model. It was already known that the two formalisms of gravity,  the metric and the Palatini formulations, lead to different inflationary predictions. In this paper we have also proven that they behave differently concerning the baryogenesis capabilities of the helical gauge fields produced at the end of inflation.

Finally there are a number of physics problems that are left open in the present work, and deserve future analysis, some of them being related to the classical problems of Higgs inflation. One of them is related to the stabilization of the Higgs potential, and the possibility of getting critical values of the Higgs mass at the inflationary scales. This problem is particularly relevant in the case where the SM flavor problem is solved by a Frogatt-Nielsen mechanism where the flavon field is coupled to the inflaton, in the way we have described in this paper. This analysis clearly requires a more detailed analysis of the renormalization group running in the presence of the Frogatt-Nielsen mechanism, working at the inflationary scales. Another obvious problem, which was outside the scope of the present paper, is the analysis of the Schwinger effect, in Higgs inflation, by numerical methods as those used in Refs.~\cite{Gorbar:2021rlt,Gorbar:2021zlr,Fujita:2022fwc,Cado:2022pxk}.

\vspace{0.5cm}
\section*{Acknowledgments}
This work is supported by the Departament d'Empresa i Coneixement, Generalitat de Catalunya Grant No.~2017SGR1069, by the Ministerio de Economía y Competitividad Grant No.~FPA2017-88915-P. IFAE is partially funded by Centres de Recerca de Catalunya. YC is supported by the European Union’s Horizon 2020 research and innovation programme under the Marie Sklodowska-Curie Actions No.~754558.

\appendix
\section{Froggatt-Nielsen mechanism in de Sitter space}
\label{sec:FN}
The Froggatt-Nielsen (FN) mechanism~\cite{Froggatt:1978nt} is one of the simplest and most elegant solutions to the problem of flavor for the SM fermions. The hierarchy of masses and mixing angles for quarks and leptons can be explained by a global, generation dependent, $U(1)$ symmetry under which the fermions are charged. This symmetry is spontaneously broken by the radial part of scalar field $S\equiv \sigma e^{i \theta}$, the ``flavon field", which is charged under the $U(1)$ (with charge conventionally normalized to -1) and which has a VEV, $\langle\sigma\rangle=v_\sigma$. The breaking is communicated to the fermion sector at different orders in the parameter $\lambda(\langle\sigma\rangle)=\langle\sigma\rangle /M_\star$,
  where $M_*$ is the scale of flavor dynamics, 
  which depend on the charges of the SM fermions $q_i$, $u^c_i$, $d^c_i$, $\ell_i$, $e^c_i$ involved in Yukawa couplings.
 
 If we denote the $U(1)$ charge of the fermion $f$ by $[f]$, the Yukawa coupling matrices are given by
 \be
 Y_u^{ij}\sim \lambda^{[q_i]+[u^c_j]},\quad Y_d^{ij}\sim \lambda^{[q_i]+[d^c_j]},\quad Y_\ell^{ij}\sim \lambda^{[\ell_i]+[e^c_j]}
 \ee
When the field $\sigma$ is at its minimum, and provided that $\lambda(v_\sigma)\simeq 0.2$, of the order of the Cabibbo angle, one can choose the $U(1)$ charges such that the SM fermion mass spectrum and mixing angles are correctly described. A simple example is provided by (see e.g.~Ref.~\cite{Babu:2009fd} for a pedagogical introduction): $[q_{3,2,1}]=[u^c_{3,2,1}]=(0,2,4)$, $d^c_{3,2,1}=(2,2,3)$, $[\ell_{3,2,1}]=(2,2,3)$, $[e^c_{3,2,1}]=(0,2,4)$. However the details of the model are not important for our argument here. 

We will introduce a coupling between the flavon and the inflaton (Higgs fields) as $|S|^2|H|^2$, and assume that the flavon field has a potential 
given, in the Jordan frame, by
\be
U(\sigma)=\lambda_1\left(|S|^2-v_\sigma^2-\lambda_2|H|^2\right)^2
\label{eq:potEW}
\ee 
which corresponds, in the Einstein frame, to the potential
\be
V(\sigma)=\frac{\lambda_1\left(\sigma^2-v_\sigma^2-\frac{1}{2}\lambda_2h^2\right)^2}{(1+\xi_h h^2/M_p^2)^2}
\ee
where $v_\sigma\gg v$, so that at electroweak scales ($h\sim v$) the vacuum expectation value $\langle\sigma\rangle \simeq v_\sigma$, which spontaneously breaks the flavor symmetry~\footnote{After the global $U(1)$ symmetry breaking a (massless) Goldstone boson will remain in the spectrum. To avoid phenomenological problems it is usually assumed that there is a small explicit soft breaking of the $U(1)$ symmetry giving a mass to the (pseudo) Goldstone boson. These model details are also orthogonal to our argument here.}. 

At the electroweak phase transition, when the field $\sigma$ is at its minimum $v_\sigma$, and provided that the flavor scale be $M_*\simeq 5 v_\sigma$, it is possible to solve the flavor problem for fermion masses. Moreover, there is an extra quartic coupling for the Higgs field from the potential (\ref{eq:potEW}) which is negligible, compared to the SM one, provided that $\lambda_1\lambda_2^2\ll \lambda_h$, where $\lambda_h$ is the SM Higgs quartic coupling evaluated at the electroweak scale. This condition can be widely satisfied e.g.~for typical values of the couplings 
\be
\lambda_1=\lambda_2=0.1
\label{eq:valores}
\ee

However during the de Sitter phase, things can be pretty much different. We will study the possibility that at the end of inflation $\lambda(\langle\sigma\rangle)\simeq 1$. In fact, at the end of inflation $h_E\simeq 10^{-2}M_p$ and one can safely neglect $v_\sigma^2$ as compared to $\frac{1}{2}\lambda_2 h_E^2$, so that $\langle\sigma\rangle\simeq \sqrt{\lambda_2/2}\,h_E$, which dictates the flavor scale $M_*$ by imposing the condition $\lambda(\langle\sigma\rangle)\simeq 1$ as 
\be
M_*\simeq \sqrt{\lambda_2/2}\,h_E\,,
\ee
which yields, e.g.~for the values of the couplings in (\ref{eq:valores}), $v_\sigma\simeq 10^{15}$ GeV.

Moreover, the condition for the de Sitter fluctuations to be suppressed, so that the field $\sigma$ stays anchored to its minimum $V(\langle\sigma\rangle)=0$, during inflation
$
V''(\langle\sigma\rangle)>\frac{9}{4}H_E^2
$~\cite{Espinosa:2015qea},
translates into the condition
\be
\frac{8\lambda_1\langle\sigma\rangle^2}{(1+\xi_h h_E^2/M_p^2)^2}>\frac{9}{4}H_E^2
\ee
which, using the value of $h_E$ above and $H_E\simeq 2\cdot 10^{13}$ GeV, yields the condition 
$\sqrt{\lambda_1\lambda_2}\gtrsim 10^{-3}$, which is satisfied for the choice in Eq.~(\ref{eq:valores}). 

%

What are the implications of the above scenario for the conductivity in the Schwinger effect? As we have seen the conductivity from a Dirac fermion $f$, of electric charge $Q_f$ and Yukawa coupling $Y_f$, is exponentially suppressed as $\sim e^{-A_f}$ where
\be
A_f=\frac{\pi Y_f^2 h^2}{2|eQ_f||E|}
\ee
 and for $A_f\gg 1$ it does not contribute to the Schwinger effect. 
 Now, considering, at the end of HI, $Y_f\sim 1$ and $h_E\simeq 10^{-2}\Mp$, the condition for the fermion $f$ to not create any conductivity, $A_f\gg 1$, self-consistently translates into an upper bound on the generated electric field $|E|$ in the absence of Schwinger effect, as
 \be
 \frac{|E|}{H_E^2}\ll \frac{10^7}{|Q_f|}
 \ee
 
The strongest bound is then provided by the leptons, for which $|Q_\ell |=1$ so that a (conservative) safe bound for all charged SM fermions to not contribute to the Schwinger effect is $E\lesssim 10^{6}H_E^2$. If we use the analytic expression for zero conductivity, $\rho_E=63/(2^{16}\pi^2\xi^3)e^{2\pi\xi}H_E^4$, we get the corresponding upper bound $\xi\lesssim 6.7$, which translates into the lower bound on the parameter $f_h$, as $f_h\gtrsim 0.0022\Mp$.

\bibliographystyle{JHEP}
\bibliography{refs}

\end{document}